\documentclass[aps,pra,twocolumn,showpacs,preprintnumbers,amsmath,amssymb,floatfix]{revtex4}
\usepackage{bbm}
\usepackage{amsmath}
\usepackage{verbatim}
\usepackage{graphicx}

\usepackage{float}
\usepackage{amssymb}
\usepackage{epsfig}

\def\<{\langle}
\def\>{\rangle}

\def\bes{\begin{displaymath}}
\def\ees{\end{displaymath}}
\def\bi{\begin{itemize}}
\def\ei{\end{itemize}}
\def\bc{\begin{center}}
\def\ec{\end{center}}
\def\bma{\begin{subequations}}
\def\ema{\end{subequations}}

\newcommand{\Hc}{{\cal H}}
\newcommand{\Ac}{{\cal A}}
\newcommand{\Bc}{{\cal B}}

\newcommand{\Tc}{{\cal T}}

\newcommand{\be}{\begin{equation}}
\newcommand{\ee}{\end{equation}}
\newcommand{\bea}{\begin{eqnarray}}
\newcommand{\eea}{\end{eqnarray}}

\newcommand{\one}{\mbox{$1 \hspace{-1.0mm}  {\bf l}$}}

\newcommand{\ad}{\ensuremath{a^\dagger}}

\begin{document}

\title{Sequential Generation of Matrix-Product States in Cavity QED}

\author{C. Sch\"on$^{1}$, K. Hammerer$^{2}$, M. M. Wolf$^{3}$,
J. I. Cirac$^{3}$, and E. Solano$^{4,5}$}

\affiliation{$^{1}$Blackett Laboratory, Imperial College London,
Prince Consort Road, London SW7 2BZ, United Kingdom\\
$^{2}$Institute for Theoretical Physics,
University of Innsbruck, 6020 Innsbruck, Austria\\
$^{3}$Max-Planck-Institut f\"ur Quantenoptik, Hans-Kopfermann-Strasse 1, 85748 Garching,
Germany\\
$^{4}$Physics Department, ASC, and CeNS,
Ludwig-Maximilians-Universit\"at, Theresienstrasse 37, 80333 Munich,
Germany \\
$^{5}$Secci\'{o}n F\'{\i}sica, Departamento de Ciencias, Pontificia
Universidad Cat\'{o}lica del Per\'{u}, Apartado Postal 1761, Lima,
Peru}

\begin{abstract}

We study the sequential generation of entangled photonic and atomic
multi-qubit states in the realm of cavity QED. We extend the work of
C. Sch\"on {\it et al.} [Phys. Rev. Lett. {\bf 95}, 110503 (2005)],
where it was shown that all states generated in a sequential manner
can be classified efficiently in terms of matrix-product states. In
particular, we consider two scenarios: photonic multi-qubit states
sequentially generated at the cavity output of a single-photon
source and atomic multi-qubit states generated by their sequential
interaction with the same cavity mode.

\end{abstract}


\maketitle

\section{Introduction}

Triggered single-photon sources
\cite{Lounis2005,Oxborrow2005,Walther2006} have important
applications in quantum communication and quantum computation and
are subject of intense experimental investigation
\cite{Kuhn2002,McKeever2004,Darquie2005,Nussmann2005,Aoki2006,Lange2004,
Blinov2004,Lounis2000,Pelton2002,Reithmaier2004,Lodahl2004,Stevenson2006,
Brouri2000,Kurtsiefer2000}. Substantial progress is being made along
various routes using single atoms
\cite{Kuhn2002,McKeever2004,Darquie2005,Nussmann2005,Aoki2006},
ions \cite{Lange2004,Blinov2004}, molecules \cite{Lounis2000},
quantum dots
\cite{Pelton2002,Reithmaier2004,Lodahl2004,Stevenson2006} or
color centers \cite{Brouri2000,Kurtsiefer2000}. The basic
operating principle common to these approaches is that a
triggered single quantum emitter excites the mode of a
cavity or a photonic crystal with a single-photon, which
coherently leaks out into a well defined field mode. If the source
is run in a cyclic fashion: sequentially initialized, loaded,
and fired, the result is ideally a train of identical single-photon
wave packets.

Given the impressive experimental achievements along these lines, it
is interesting to go one step further and address the
following scenario. Assume that the source is not initialized
after each step, but stays in some quantum state, which can in
turn be correlated to the field state generated so far. What kind
of multipartite quantum states can then in principle be created with such a
sequential generation scheme? We answered this question in Ref.~\cite{Schoen2005} and proved that the
class of sequentially generated states is exactly identical to a
class of so-called matrix-product-states (MPS)
\cite{David,Klumper,Fannes1992}. These states play an important
role in a completely different context, namely in the theory of
one-dimensional spin chains \cite{Affleck1988}, where they
constitute the set of variational states over which Density Matrix
Renormalization Group techniques are carried out
\cite{Ostlund1995,VidalDMRG,FrankDMRG}. This classification is significant
in at least three respects: (i) It does not only apply to all the
various types of single-photon sources
\cite{Lounis2005,Oxborrow2005,Walther2006,
Kuhn2002,McKeever2004,Darquie2005,Nussmann2005,Aoki2006,Lange2004,
Blinov2004,Lounis2000,Pelton2002,Reithmaier2004,Lodahl2004,Stevenson2006,
Brouri2000,Kurtsiefer2000}, but literally to any system where a
multipartite quantum state is generated by sequential interaction
with a source- or ancilla-system~\cite{Delgado2006}. (ii) It
provides a constructive protocol for the generation of any desired
quantum state. It allows one to decide whether or not a given
state can be generated with a given setup and, if so, what sequential
operations one has to apply. (iii) It stresses the
importance of the MPS formalism as it demonstrates that they naturally occur in
vastly different physical contexts.

In this work, we elaborate on the results of Ref.~\cite{Schoen2005},
providing more details and further applications. In section
\ref{sec:SequGen}, we first illustrate the basic idea of sequential
generation of quantum states by means of a cavity QED (CQED)
single-photon source \cite{Kuhn2002, McKeever2004, Lange2004}. Then,
we give a detailed derivation and extension of the main results in
Ref.~\cite{Schoen2005} without reference to any specific setup. In
section \ref{sec:PhotonicStates}, we show how to create important
entangled multi-qubit photonic states encoded in photonic time bins
and in polarization states. In section \ref{sec:AtomicStates}, we
consider a microwave cavity interacting with a sequence of atoms
flying through it, such as in Refs.~\cite{Haroche2000,Walther2001},
generating entangled multi qubit atomic states.

\section{Sequential generation of entangled multi-qubit
states}\label{sec:SequGen}

In this section, we consider first the CQED single-photon source
\cite{Kuhn2002, McKeever2004, Lange2004} and show, with an elementary
example, how the structure of MPS arises quite naturally in this
setup. We then treat the sequential generation scenario without
referring to any particular physical system. First, we assume that
an arbitrary source-qubit interaction is available in each step of
the sequential generation. In the second part, we restrict the
interaction in a way which resembles the situation in current
cavity QED setups for the generation of single-photon pulses.
Finally, we discuss a more abstract scenario, a register of
qubits, to which nearest neighbor gates are applied sequentially.
This situation is again covered conveniently by the MPS formalism,
even if the sequence of gates is applied repeatedly.

\subsection{Basic idea}

\begin{figure}[t]
\begin{center}
\epsfig{file=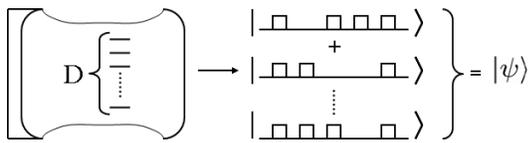,angle=0,width=0.8\linewidth}
\end{center}
\caption{A trapped $D$-level atom is coupled to a cavity qubit,
determined by the energy eigenstates $|0\>$ and $|1\>$. After
bipartite source-qubit operations, photonic time-bins are
sequentially and coherently emitted at the cavity output, creating a
desired entangled multi-qubit stream.} \label{source}
\end{figure}

The CQED single-photon source utilizes the possibility to excite the
cavity mode via an atom, which is trapped inside the
cavity~\cite{Kuhn2002, McKeever2004,Lange2004}. The photon is then
coherently emitted through the cavity mirror. A photonic qubit may
be defined either by different polarization states or by the absence
and the presence of the photon. If we allow for specific operations
inside the source before each photon emission, we will be able to
create different multi-qubit states at the output (see Fig.~\ref{source}).

Typically, a CQED single-photon source employs an effective two-level
atom with hyperfine ground states $|a\>$ and $|b\>$ coupled via a
Raman transition driven by the cavity mode and an external laser.
The latter controls the population transfer \be |a,0\> \rightarrow
\cos(\phi_1) |a,0\> + \sin(\phi_1) e^{i\varphi_1} |b,1\>,\ee where
$|0\>$ and $|1\>$ denote the absence and the presence of a photon in
the cavity mode. This so-called time-bin qubit is then coherently
emitted as depicted in Fig.\ \ref{source}.

\begin{figure}[t!]
\begin{center}
\epsfig{file=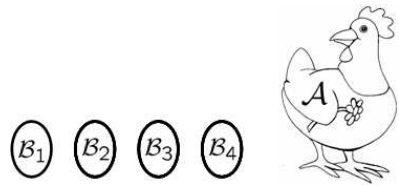,angle=0,width=0.6\linewidth}
\end{center}
\caption{The hen and egg picture: The ancilla $\Ac$ couples
sequentially to initially uncorrelated qubits $\Bc_i$. In the last
step we require the ancilla to decouple from the entangled $n$-qubit
state.} \label{henegg}
\end{figure}

We define $c_i= \cos(\phi_i)$ and $s_i= \sin(\phi_i) e^{i\varphi_i}$
for step $i$ and repeat the procedure $n$ times. We end up with
\bea |a\> &\rightarrow & c_1 |a,0\> + s_1 |b,1\> \nonumber\\
&\rightarrow & c_1 c_2 |a,0,0\> + |b\> \big( c_1 s_2 |1,0\> + s_1
|0,1\>\big) \nonumber\\ &\rightarrow & \dots \nonumber\\
&\rightarrow & |b\> \big[c_1 \dots c_{n-1} |1,0,\dots,0\>
\nonumber\\&& +\;c_1 \dots c_{n-2} s_{n-1} |0,1,0,\dots,0\> + \dots
\nonumber\\&& +\;c_1 s_2 |0,\dots,0,1,0\> + s_1 |0,\dots,0,1\>
\big], \eea where we chose $c_n=0$ and $s_n=1$. Then, the atom
decouples and the resulting photonic state is a W-type $n$-qubit
state. This is just a particular example of a more general sequential
generation scheme (see Fig.~\ref{henegg}).

An ancillary system $\Ac$ (the atom) with Hilbert space
$\Hc_\Ac\simeq\mathbb{C}^D$ (D=2) couples sequentially to initially
uncorrelated qubits $\Bc_i$ (the time-bin qubits) with Hilbert
spaces $\Hc_\Bc\simeq\mathbb{C}^2$. In every step we have a unitary
time evolution of the joint system $\Hc_\Ac\otimes\Hc_\Bc$. Since
each qubit is initially in the state $|0\>$, we disregard the qubits
at the input and write the evolution in the form of an isometry
$V:\Hc_\Ac\rightarrow\Hc_\Ac\otimes\Hc_\Bc$. The photon-generating
process discussed above is given by \be \label{V_intro} V_{[i]} =
c_i |0\>_i|a\>\<a| + s_i |1\>_i|b\>\<a| + |0\>_i|b\>\<b|,\ee for
step $i$ and fulfills the isometry condition
$V_{[i]}^{\dagger}V_{[i]}= \one_2$. The final state can then be
written as \be |\Psi\> = V_{[n]}\dots V_{[1]} |\varphi_I\>, \ee
where $|\varphi_I\>=|a\>$ is the initial state of the atom. Since
the atom decouples in the last step, we may also write \bea |\Psi\>
&=& |\varphi_F\> \<\varphi_F| V_{[n]}\dots V_{[1]} |\varphi_I\>
\nonumber\\ &=& |\varphi_F\> |\psi\>, \eea where $|\varphi_F\>=|b\>$
is the final state of the atom and $|\psi\>$ the $n$-qubit photonic
state. The goal in the following sections is to classify all
achievable states $|\psi\>$ in terms of the required resources such
as number of ancilla levels $D$ and possible operations on the
atom-cavity system $V_{[i]}$.

\subsection{Arbitrary source-qubit interaction}

\begin{figure}[t]
\begin{center}
\epsfig{file=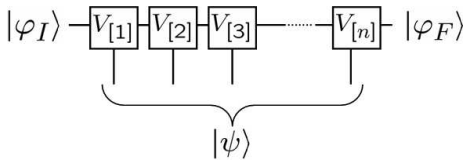,angle=0,width=0.7\linewidth}
\end{center}
\caption{Sequential generation of a multi-qubit state $|\psi\>$. In
each step the $D$-dimensional ancilla produces one qubit. This
process is described by a $2D \times D$ dimensional isometry
$V_{[i]}$. The achievable multi-qubit states are instances of
matrix-product states with $D$ dimensional bonds and open boundary
conditions specified by $|\varphi_I\>$ and $|\varphi_F\>$.}
\label{fig:ancilla}
\end{figure}

Now, we want to look at the problem from a more general perspective.
We assume that the operators $V_{[i]}$ are arbitrary isometries and
that the ancilla decouples in the last step. We express the
isometries in a given basis \be
V=\sum_{i,\alpha,\beta}V_{\alpha,\beta}^i |\alpha, i \>\<\beta|, \ee
where each $V^i$ is a $D \times D$ matrix and $\{ | \alpha \rangle ,
| \beta \rangle \}$ are any of the $D$ ancillary levels. This is the
generalization of Eq.~(\ref{V_intro}) and the isometry condition
then reads \be V^{\dagger} V = \sum_{i=0}^1 V^{i\dagger} V^i
=\one_D.\ee  The resulting $n$-qubit state is then given by
\be\label{eq:MPSiso} |\psi\>= \sum_{i_1\ldots i_n=0}^1\<\varphi_F|
V_{[n]}^{i_n}\ldots V_{[1]}^{i_1}|\varphi_I\>\;|i_n,\ldots, i_1\>.
\ee This is a matrix-product state (MPS)~\cite{Klumper,Fannes1992}
with $D$ dimensional bonds and open boundary conditions, which are
specified by the initial ancilla state $|\varphi_I\>$ and the final
ancilla state $|\varphi_F\>$. In Figs.~\ref{henegg} and
\ref{fig:ancilla}, we illustrate the sequential generation process
schematically.

In practice, the question whether a given state $|\tilde{\psi}\>$
can be generated with certain resources is more important.
Therefore we have to show that every MPS of the form
\be\label{MPS} |\tilde{\psi}\> = \<\tilde{\varphi}_F|
\tilde{V}_{[n]}\ldots \tilde{V}_{[1]}|\tilde{\varphi}_I\>,\ee with
arbitrary maps
$\tilde{V}_{[k]}:\Hc_\Ac\rightarrow\Hc_\Ac\otimes\Hc_\Bc$, can be
generated by isometries of the same dimension such that the
ancilla decouples in the last step. Since the idea of the proof is
an explicit construction of all involved isometries, it will
provide a general recipe for the sequential generation of
$|\tilde{\psi}\>$. Note that every state has a MPS representation
\cite{David,Vidal2003}.

We start by writing \be \big(\<\tilde{\varphi}_F|\otimes\one_2\big)
\tilde{V}_{[n]}= V_{[n]}'M_{[n]}, \ee where the $2\times 2$ matrix
$V_{[n]}'$ is the left unitary in the singular value decompositions
(SVD) of the left hand side and $M_{[n]}$ is the remaining part.

The recipe for constructing the isometries is the induction \be
\big(M_{[k]}\otimes\one_2\big)\tilde{V}_{[k-1]}
=V_{[k-1]}'M_{[k-1]}\label{induction},\ee where the isometry
$V_{[k-1]}'$ is constructed from the SVD of the left hand side, and
$M_{[k-1]}$ is always chosen to be the remaining part.

After $n$ applications of Eq.~(\ref{induction}) in Eq.~(\ref{MPS}),
from left to right, we set
$|\varphi_I\>=M_{[1]}|\tilde{\varphi}_I\>$, producing \be
|\tilde{\psi}\> = V_{[n]}'\ldots V_{[1]}'|\varphi_I\>.\ee Simple
rank considerations show that $V_{[n-k]}'$ has dimension
$2\min{[D,2^k]}\times \min{[D,2^{k+1}]}$. The dimension of the left
unitary grows exponentially, i.e. $V'_{[n-k+1]}$ has dimension $2^k
\times 2^k$, as long as $2^k< D$. For $2^{k+1}>D$, superfluous
columns appear in $V'_{[n-k]}$ since the original matrix
$\big(M_{[n-k+1]}\otimes\one_2\big) \tilde{V}_{[n-k]}$ has at most
$D$ singular values. Truncation leads to a $2^{k+1} \times D$
dimensional isometry $V'_{[n-k]}$.

Now $M_{[n-k]}$ has dimension $D\times D$ and all subsequent left
isometries have dimension $2D \times D$. Therefore every $V_{[k]}'$
can be embedded into an isometry $V_{[k]}$ of dimension $2D \times
D$. Physically, this means that we have redundant ancillary levels
which we do not use. Finally, decoupling the ancilla in the last
step is guaranteed by the fact that, after the application of
$V_{[n-1]}$, merely two levels of $\Hc_\Ac$ are yet occupied, and
can be mapped entirely onto the system $\Hc_\Bc$. This is precisely
the action of $V_{[n]}$ through its embedded unitary $V_{[n]}'$.

Together with Eqs.\ (\ref{eq:MPSiso}) and (\ref{MPS}), this proves the
equivalence of three sets of $n$-qubit states:
\begin{enumerate}
    \item MPS with $D$-dimensional bonds and open boundary
    conditions.
    \item States which are generated
sequentially and isometrically by a $D$-dimensional ancillary
system which decouples in the last step. That is, the generation
is deterministic.
    \item States which can be generated sequentially by a
    $D$-dimensional ancillary system in a probabilistic manner.
    That is, the preparation may only be successful with some probability
    and include measurements and conditional
    operations.
\end{enumerate}
In Secs. \ref{subsec:standard} and \ref{subsec:directinteraction}, we
will show two other equivalent classes. We emphasize that this
result  holds as well for higher dimensional systems (beyond
qubits) and that its constructive proof provides a recipe for the
sequential generation of any state with minimal resources, namely
a $D$-dimensional source/ancilla for a $D$-dimensional MPS.

Note that in order to obtain MPS with periodic boundary conditions
we require an interaction between the first and the last qubit.
For systems, where an ancilla is available, one may store the
first qubit within the ancilla. Then, in order to produce MPS with
$D$-dimensional bonds, we would require a $2D$ dimensional system
to store the additional qubit. On the other hand many interesting
states belong to the class of MPS with $2$-dimensional bonds and
in this case two additional atomic levels would suffice.

\subsection{The standard map}\label{subsec:standard}

The situation considered above assumes that arbitrary isometries
can be achieved in the generation of a qubit, which amounts to
have complete control over the source-qubit interaction. This does
not quite correspond to what is the case in current cavity QED
single-photon sources \cite{Kuhn2002, McKeever2004, Lange2004},
where only atomic degrees of freedom can be manipulated easily,
while the isometry describing the generation of a photon (qubit)
is fixed. In the following we will show that also in this
restricted scenario it is possible to generate arbitrary MPS with
$D$-dimensional bonds if a $2D$-level atom is used as a source.

\begin{figure}[t]
\begin{center}
\epsfig{file=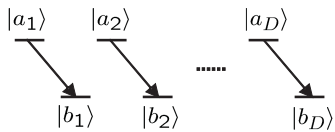,angle=0,width=0.5\linewidth}
\end{center}
\vspace*{-0.5cm}\caption {Restricted interaction between ancilla
and qubit. Each atomic transition from $|a_i\>$ to its
respective $| b_i \>$ is accompanied by the generation of a photon
in a certain time-bin.} \label{std}
\end{figure}

We consider an atomic system with $D$ states $| a_i \rangle$ and $D$
states $| b_i \rangle$, as depicted in Fig.\ \ref{std}, so that
$\Hc_\Ac=\Hc_a\oplus\Hc_b\simeq \mathbb{C}^D\otimes\mathbb{C}^2$.

That is, we will write $|\varphi\>|1\>$ for a superposition of
$|a_i\>$ states, whereas $|\varphi\>|0\>$ denotes a superposition of
$|b_i\>$ states. Since the last qubit marks whether the atomic level
belongs to the $|a_i\>$ or to the $| b_i \rangle$ subspace, we will
refer to it as the tag-qubit and write
$\Hc_\Ac=\Hc_{\Ac'}\otimes\Hc_\Tc$.

\begin{figure}[ht]
\begin{center}
\epsfig{file=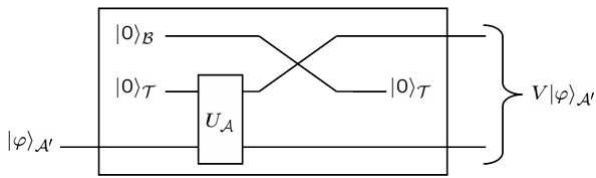,angle=0,width=0.9\linewidth}
\end{center}
\vspace*{-0.5cm}\caption {The box represents a single step in the
sequential generation scheme. Since tag- and time-bin qubit are
always in state $|0\>$ after the preceding step, they can be ignored
at the input. After an arbitrary unitary operation $U_{\Ac}$ on the
atom, tag- and time-bin qubit are interchanged by the standard map.
Ignoring the tag qubit also at the output (since it is always in
$|0\>$), the process can be described by an arbitrary isometry
$V$ applied on the effective ancillary system $\Ac'$.} \label{step}
\end{figure}

Now assume atomic transitions from each $|a_i\>$ state to its
respective $|b_i\>$ state are accompanied by the generation of a
photon in a certain time-bin. This is described by a unitary
evolution, from now on called ``D-standard map'', of the form \bea
T:\;|\varphi\>_{\Ac'}|1\>_\Tc |0\>_\Bc &\mapsto&
|\varphi\>_{\Ac'}|0\>_\Tc |1\>_\Bc\;, \nonumber
\\ |\varphi\>_{\Ac'}|0\>_\Tc |0\>_\Bc &\mapsto&
|\varphi\>_{\Ac'}|0\>_\Tc |0\>_\Bc\; . \eea  Hence, $T$ effectively
interchanges the tag-qubit with the time-bin qubit. If,
additionally, arbitrary atomic unitaries $U_{\Ac}$ are allowed at
any time, we can exploit the swap caused by $T$ in order to generate
the operation \be V|\varphi\>=\<0|_\Tc
T\Big(U_{\Ac}\big(|\varphi\>_{\Ac'}|0\>_\Tc\big)|0\>_\Bc\Big)
\label{IsofromT},\ee which is the most general isometry
$V:\Hc_{\Ac'}\rightarrow \Hc_{\Ac'}\otimes\Hc_\Bc$. Equation
(\ref{IsofromT}) is also illustrated in Fig.\ \ref{step}.

Therefore, the so generated $n$-qubit states include all possible
states arising from subsequent applications of $2D \times
D$-dimensional isometries. On the other hand, they are a subset of
the MPS in Eq.~(\ref{MPS}) with arbitrary $2D \times
D$-dimensional maps, assuming that the atom decouples at the end.
Hence, this set is again equivalent to the three mentioned above.

\subsection{Qubit-qubit interaction without
ancilla}\label{subsec:directinteraction}

\begin{figure}
\begin{center}
\epsfig{file=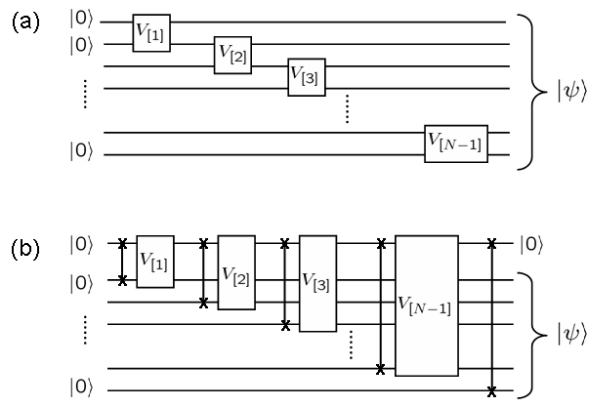,angle=0,width=0.9\linewidth}
\end{center}
\vspace*{-0.5cm}\caption {(a) Two-qubit gates $V_{[i]}$ are
sequentially applied between nearest neighbor qubits. This situation
can be simulated by a two-dimensional ancilla as demonstrated in
(b). Instead of applying the gate between neighboring qubits $k$ and
$k+1$, we use qubit $k$ and the ancilla and swap the ancilla state
afterwards with the qubit $k+1$. Since we can merge the swap
operations and the arbitrary unitary operations $V_{[i]}$ to
arbitrary unitary operations, we know that the class of achievable
states $|\psi\>$ is equivalent to the class of MPS with open
boundary conditions and two-dimensional bonds.} \label{direct}
\end{figure}

In Fig.\ \ref{direct} (a) we depicted a system of $N$ initially
uncorrelated qubits, which interact sequentially with their nearest neighbor
qubit.

This situation is in fact identical to the one considered so far,
as one can imagine the operation $V_{[k]}$ being performed not
between qubit $k$ and $k+1$ directly but between qubit $k$ and a
two-dimensional ancilla, which is then swapped on qubit $k+1$ [see
Fig.\ \ref{direct} (b)]. In the last step, the swap ensures that
the ancilla decouples from the desired multi-qubit state
$|\psi\>$. Thus, also for direct qubit-qubit interaction, the class
of achievable states $|\psi\>$ is equivalent to the class of MPS
with two-dimensional bonds (see \cite{David} for a more formal
argument). In section \ref{sec:AtomicStates}, we will come back to
this scenario and consider an experimental setup, where always two
neighbor atoms of a chain interact via a common cavity mode.

\begin{figure}
\begin{center}
\epsfig{file=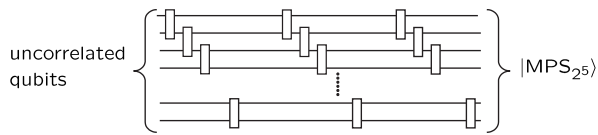,angle=0,width=0.9\linewidth}
\end{center}
\vspace*{-0.5cm}\caption {Repeating the procedure $m$ (here $m=3$)
times defines a class of states described by MPS with
$2^{2m-1}$-dimensional bonds. Note that in between the three
sequences we also allow an arbitrary interaction between the first
and the last qubit.} \label{repeat}
\end{figure}

If direct two-qubit interaction between neighbors is possible, there
exists at least in principle -- in contrast to the hen and egg
scenario -- no reason why one should not apply them more than once.
If this is done $m$ times, as illustrated in Fig.\ \ref{repeat},
still in a sequential manner, i.e. after an operation between the
first and the last qubit one starts again with qubits $1$ and $2$,
the achievable class of states will be described by MPS with bonds
of dimension $d^{2 m-1}$, that is for qubits $2^{2m-1}$. However, in
this case the two sets will no longer be equivalent, i.e., there are
MPS with $D=2^{2m-1}$ for whose preparation we need more than $m$
such layers of two-qubit gates.

\section{Sequential generation of photonic
states}\label{sec:PhotonicStates}

There are two possible ways of encoding quantum information in
photon pulses: orthogonal polarisation states and energy
eigenstates. While the latter is more straightforward in terms of
the required resources, polarization encoding \cite{Gheri2000,
Wilk2006} avoids the trouble caused by a failure of the source for
the generation of multi-qubit entangled states since a missing
photon is not being mistaken as an empty time-bin, i.e., the success
of the encoding step is heralded by the observation of a photon.

The generation of multi-photon entangled states using several
distant single photon sources was explored in Ref. \cite{Beige2006}.

\subsection{Time-bin entanglement}

We will first consider a situation where cavity decay is negligible
on the time-scale of the operations performed on the atom-cavity
system. After the cavity qubit has left, we start with the next step
and repeating the process leads to a multi-qubit entangled photon
state at the output as sketched in Fig.~\ref{source}.

\subsubsection{Arbitrary source-qubit operation}

\begin{figure}[ht]
\begin{center}
\epsfig{file=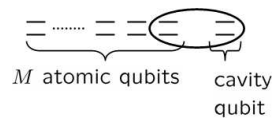,angle=0,width=0.4\linewidth}
\end{center}
\caption{The $D$-level atom can be viewed as a set of $M$ qubits
with $D\le 2^M$. For an arbitrary operation we need local unitaries
for the atomic qubits and a universal two-qubit gate between the
cavity qubit and one specific atomic qubit.} \label{fig:levels}
\end{figure}

In the following, we demonstrate how an arbitrary unitary operation
on the $2D$ dimensional Hilbert space of the combined atom-cavity
system can be realized. Therefore, we view the $D$-level atom as a
set of $M$ qubits with $D\le 2^M$ as depicted in
Fig.~\ref{fig:levels}. Then, we have to show that one can perform
arbitrary two-qubit operations between each pair of qubits, i.e.,
universal quantum computing. Since the atomic levels can be
manipulated at will using Raman laser systems~\cite{NIST,Innsbruck},
it remains to propose an arbitrary two-qubit gate between one
specific atomic qubit and the cavity qubit.

\begin{figure}[t]
\begin{center}
\epsfig{file=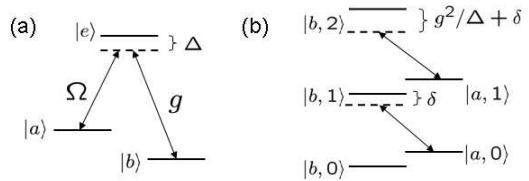,angle=0,width=0.8\linewidth}
\end{center}
\caption{(a) Atomic level structure: levels $|a\rangle$ ($|b\>$) and
$|e\>$ are coupled by a laser (cavity mode) off resonance. (b) After
adiabatic elimination of the upper state $|e\>$, we are left with a
Jaynes-Cummings type of Hamiltonian, where states $|a,n\>$ and
$|b,n+1\>$ are coupled. Both, the energy difference of those levels
and the corresponding Rabi frequency depends on $n$. The reason for
the first is the ac-Stark shift, whereas the second is due to the
Jaynes-Cummings coupling.} \label{sel}
\end{figure}

Therefore, we consider a typical three-level lambda configuration
(see Fig.~\ref{sel}), where the hyperfine ground states $|a\>$ and
$|b\>$ are coupled to the excited level $|e\>$ off-resonantly
through a laser with Rabi frequency $\Omega$ and detuning $\Delta+
\delta$, and the cavity mode $a$ with coupling strength $g$ and
detuning $\delta$.

Furthermore, we assume that the cavity decay rate $\kappa$ is smaller
than any other frequency in the problem, so that we can ignore
cavity damping during the atom-cavity manipulations. In an
appropriate interaction picture, the Hamiltonian of the system is
then given by \bea H&=& -\Delta\,\big(\sigma_{aa}+\ad a\big)+
g\big(\sigma_{eb}a+ \ad\sigma_{be}\big) \nonumber\\ &&+
\frac{\Omega}{2} \big(e^{-i\delta t}\sigma_{ea}+e^{i\delta
t}\sigma_{ae}\big),\eea with $\sigma_{kl}=|k\>\<l|$,
$\{k,l\}=\{a,b,e\}$ and $|\Delta|\gg g,\Omega \gg \delta$. After
adiabatically eliminating the excited state $|e\>$, the Hamiltonian
for the effective $D=2$ atomic system plus cavity mode, in an
interaction picture with respect to $-\Delta(\sigma_{aa}+\ad a)$, is
given by \be\label{Hadia} H_{\rm ad}=
\frac{\Omega^2}{4\Delta}\;\sigma_{aa}+ \frac{g^2}{\Delta}\;\ad
a\,\sigma_{bb}+ \frac{g\Omega}{2\Delta}\big(e^{-i\delta t}
\sigma_{ab}a+ e^{i\delta t} \ad \sigma_{ba}\big).\ee It describes an
effective Jaynes-Cummings coupling between the cavity mode and the
atomic $|a\> \to |b\>$ transition with Rabi frequency
$g\Omega/2\Delta$. The other terms correspond to ac-Stark shifts. In
an interaction picture with respect to the latter the Hamiltonian is
given by \be H_{\rm ad}^{I}= \sum_{n=0}^{\infty}
\frac{\sqrt{n}g\Omega}{2\Delta} \left(e^{-i \big(\delta-
\frac{\Omega^2}{4\Delta} +\frac{ng^2}{\Delta}\big)t}
|a,n-1\>\<b,n|+ {\rm H.c.} \right). \ee

Then, we choose the laser frequency such that \be \delta=
\frac{\Omega^2}{4\Delta} -\frac{g^2}{\Delta}.\ee For $g^2/\Delta \gg
g\Omega/2\Delta$, the effective interaction in all subspaces with $n
\neq 0$ is then dispersive and the selective
Hamiltonian~\cite{FrancaSantos2001} is given by \bea H_{\rm sel} &=&
\frac{g\Omega}{2 \Delta} \big( |a,0\>\<b,1|+ |b,1\>\<a,0|
\big)\nonumber\\ &=& \frac{g\Omega}{4\Delta} \big(\sigma_x^{\Ac}
\otimes \sigma_x^{\Bc}+ \sigma_y^{\Ac} \otimes \sigma_y^{\Bc}
\big),\label{eq:hsel}\eea where $\sigma_i^{\Ac}$ and
$\sigma_i^{\Bc}$ denote the Pauli matrices acting on the atomic and
the photonic qubit, respectively. Using a laser pulse of an
appropriate duration, we obtain the entangling two-qubit gate
\be\label{eq:iswap} \sqrt{\rm ISWAP}= \exp\big[i\pi\big( |a,0\>\<
b,1| + |b,1\>\< a,0|\big)/4\big].\ee Together with local operations
in both qubits~\cite{FrancaSantos2005}, this suffices to generate an
arbitrary two-qubit operation.

\subsubsection{Adiabatic passage}

In current CQED single-photon sources \cite{Kuhn2002, McKeever2004,
Lange2004, Darquie2005}, an adiabatic passage is employed to realize
the $1$-standard map introduced above. Using one additional level,
we will show how to generate familiar multi-qubit states like
W~\cite{Wstates}, GHZ~\cite{GHZ}, and cluster
states~\cite{Briegel2001}, which are all MPS with $D=2$
\cite{VerstCirQC}.

For this purpose, we consider an atom with three effective levels
$\{|a\>,|b_1\>,|b_2\>\}$ trapped inside an optical cavity. With the
help of a laser beam, state $|a\rangle$ is mapped to the internal
state $|b_1\>$, and a photon is generated, whereas the other states
remain unchanged. This physical process is described by the map \bea
M_{\Ac\Bc}:\; |a\>  &\mapsto& |b_1\>|1\>, \nonumber\\
|b_1\> &\mapsto& |b_1\>|0\>, \nonumber\\ |b_2\> &\mapsto&
|b_2\>|0\>,\eea and can be realized with the techniques used
in~Refs.~\cite{Kuhn2002,McKeever2004,Lange2004}. After the
application of this process, an arbitrary operation is applied to
the atom, which can be performed with Raman transitions. The
photonic states that are generated after several applications are
those MPS with isometries $V_{[i]}=M_{\Ac\Bc}U_\Ac^{[i]}$, with
${i=1,\dots,n}$, $U_\Ac^{[i]}$ being arbitrary unitary atomic
operators.

For example, to generate a W-type state of the form \bea |\psi_{\rm
W}\> &=& e^{i\Phi_1}\sin\Theta_1 |0...01\> + \cos\Theta_1
e^{i\Phi_2}\sin\Theta_2 |0...010\> \nonumber\\&& +...+
\cos\Theta_1...\cos\Theta_{n-2} e^{i\Phi_{n-1}}\sin\Theta_{n-1}
|010...0\>\nonumber\\&& + \cos\Theta_1...\cos\Theta_{n-1} |10...0\>,
\eea we choose the initial atomic state $|\varphi_I\>=| b_2 \>$ and
operations $U_\Ac^{[i]} = U_{a {b_2}}^{b_1}(\Phi_i,\Theta_i)$, with
$i=1,\dots,n-1$, where \bea U_{kl}^m(\Phi_i,\Theta_i) & \!\!\!\!\! =
\!\!\!\! & \cos\Theta_i|k\>\<k| + \cos\Theta_i |l\>\<l|+ e^{i\Phi_i}
\! \sin\Theta_i|k\>\<l| \nonumber\\&& - e^{-i\Phi_i} \sin\Theta_i
|l\>\<k| + |m\>\<m| ,\eea and $\{ k,l,m \} = \{ a,b_1,b_2 \}$. To
decouple the atom from the photon state, we choose the last atomic
operation $U_\Ac^{[n]} = U_{a {b_2}}^{b_1}(0,\pi/2)$ and, after the
last map $M_{\Ac\Bc}$, the decoupled atom will be in state $|b_1\>$.

To produce a GHZ-type state in similar way, we choose $|\varphi_I\>=
|a\>$, $U_\Ac^{[1]}=U_{a {b_2}}^{b_1}(\Phi_1,\Theta_1)$,
$U_\Ac^{[i]} = U_{a {b_1}}^{b_2}(0,\pi/2)$, with $i=2,\dots,n-1$,
and $U_\Ac^{[n]} = U_{{b_1} {b_2}}^a(0,\pi/2)U_{a
{b_1}}^{b_2}(0,\pi/2)$.

For generating cluster states, we choose $|\varphi_I\>=|b_2\>$,
$U_\Ac^{[i]} = U_{a {b_2}}^{b_1}(\Phi_i,\Theta_i) U_{a
{b_1}}^{b_2}(0,\pi/2)$, with $i=1,\dots,n-1$, and $U_\Ac^{[n]} =
U_{a {b_1}}^{b_2}(\Phi_n,\Theta_n) U_{{b_1}{b_2}}^a(0,\pi/2) U_{a
{b_1}}^{b_2}(0,\pi/2)$, obtaining \be |\psi \> = \bigotimes_{i=1}^n
\left(O_{i-1}^0 |0\>_i + O_{i-1}^1 |1\>_i\right) , \ee where
$O^0_{i-1} = \cos\Theta_i |0\>_{i-1}\<0| -e^{-i\Phi_i}\sin\Theta_i
|1\>_{i-1}\<1|$ and $O^1_{i-1}= e^{i\Phi_i}\sin\Theta_i
|0\>_{i-1}\<0| + \cos\Theta_i |1\>_{i-1}\<1|$, with
${i=2,\dots,n-1}$. Operators $O_{i-1}^0$ and $O_{i-1}^1$ act on the
nearest neighbor-qubit $i-1$ under the assumption $O_0^0 \equiv
\cos\Theta_1$ and $O_0^1 \equiv e^{i\Phi_1}\sin\Theta_1$. If one
chooses $\Phi_i=0$ and $\Theta_i=\pi/4$, this leads to the cluster
states defined by \be\label{cluster} |\psi_{\rm cl}\> =
\frac{1}{2^{n/2}} \bigotimes_{i=1}^n \left(\sigma^z_{i-1}|0\>_i +
|1\>_i\right) \,\, , \,\, {\rm with} \,\,\, \sigma^z_0\equiv 1 . \ee

\subsection{Polarization entanglement}

Here, the cavity qubit is defined by single excitations in two modes
$a$ and $b$, which have equal frequencies but orthogonal
polarizations. As above for the time-bin qubits, we will first show
how to realize arbitrary source-qubit operations and then focus on
the less demanding scenario, where the photonic qubits are generated
by a standard map, i.e., an adiabatic passage.

\subsubsection{Arbitrary source-qubit operation}

\begin{figure}[t]
\begin{center}
\epsfig{file=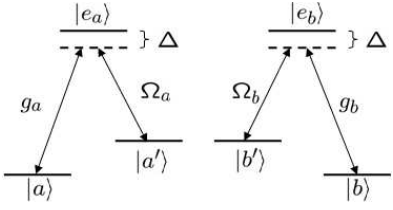,angle=0,width=0.6\linewidth}
\end{center}
\vspace*{-0.5cm}\caption {Levels $|a'\>$ and
$|b'\>$ ($|a\>$ and $|b\>$) are coupled off-resonance to $|e_a\>$ and $|e_b\>$ by
two lasers (cavity modes).} \label{polsel}
\end{figure}

In this section, we will suggest how to implement an arbitrary
operation on the atom-cavity system based on present cavity QED
experiments \cite{Kuhn2002,McKeever2004,Lange2004}. Therefore, one
has to show how an arbitrary unitary operation on one specific
atomic qubit $\{|a\>, |b\>\}$ and the cavity qubit $\{|1_a\>,
|1_b\>\}$ can be realized. We consider a double-lambda type atomic
level configuration as illustrated in Fig.\ \ref{polsel}, which
couples to the two cavity modes via two Raman transitions.

Two external laser fields drive the transitions from level $|a'\>$
to the excited level $|e_a\>$ and from level $|b'\>$ to the excited
level $|e_b\>$ with Rabi frequencies $\Omega_a$ and $\Omega_b$,
respectively. The cavity modes $a$ and $b$ couple to the transitions
between $|e_a\>$ and level $|a\>$ with coupling strength $g_a$ and
$|e_b\>$ and level $|b\>$ with coupling strength $g_b$.

For large detunings \be\label{pse_deltabig} |\Delta_a|, |\Delta_b|\gg
g_a,g_b,\Omega_a,\Omega_b \ee we can adiabatically eliminate the
excited levels $|e_a\>$ and $|e_b\>$ and end up with a
Jaynes-Cummings type of Hamiltonian, which is block separable in the
subspaces spanned by $\{|a',n_a\>,|a,n_a+1_a\>\}$ and
$\{|b',n_b\>,|b,n_b+1_b\>\}$, where $n_a$ and $n_b$ denote the
number of photons in mode $a$ and $b$. As above, the ac-Stark shifts
for $n_a=n_b=0$ can be compensated by choosing the frequency of the
laser fields appropriately. Under the condition \be \frac{n_ag_a^2}{\Delta_a} \gg
\frac{g_a\Omega_a}{2\Delta_a},~ \frac{n_bg_b^2}{\Delta_b} \gg
\frac{g_b\Omega_b}{2\Delta_b}, \ee we obtain the selective
Hamiltonian \be H_{\rm sel}^p= \frac{g_a\Omega_a}{2\Delta_a}
|a',0_a\>\<a,1_a| +\frac{g_b\Omega_b}{2\Delta_b} |b',0_b\>\<b,1_b|
+{\rm H.c.},\ee where $|0_a\>$ and $|0_b\>$ denote the empty cavity
modes $a$ and $b$. For appropriate $\Omega_a(t)$ and $\Omega_b(t)$,
the evolution operator reads \be U_{\rm sel}^p= \exp\big[i\pi \big( |a',0_a\>\<a,1_a|+
|b',0_b\>\<b,1_b|+ {\rm H.c.} \big)/2\big].\ee

After the $k$th photonic qubit leaked out of the cavity, the state
of the system is given by \be |\Psi_k\>= \alpha |a\> |\psi_k^a\>+
\beta |b\> |\psi_k^b\>,\ee where $|\psi_k^a\>$ and $|\psi_k^b\>$ are
$k$-qubit photonic states. Now we have to initialize the system for
the next step, i.e., provide the polarization qubit \bea |\Psi_k\>
&\rightarrow& -i\big(\alpha
|a',0_a\> |\psi_k^a\> + \beta |b',0_b\> |\psi_k^b\> \big) \nonumber\\
&\rightarrow& \alpha |a,1_a\> |\psi_k^a\> + \beta |b,1_b\>
|\psi_k^b\>,\eea where we applied $U_{\rm sel}^p$ in the second
line. In order to realize an arbitrary two-qubit gate, we combine
local operations with a $\sqrt{\rm ISWAP}$ two-qubit gate between
the atomic and the photonic qubit. The latter can be achieved in
three steps, \be \sqrt{\rm ISWAP}= \left(U_{\rm sel}^p\right)^{-1}
e^{i\pi(|a'\>\<b'|+|b'\>\<a'|)/4}~ U_{\rm sel}^p.\ee

It remains to show how to decouple the atom from the generated
multi-photon state in the final step $n$. Therefore, one has to map
the atomic state on the last photonic qubit: after the
transformation of the atomic levels $|a\>\rightarrow -i|a'\>$ and
$|b\>\rightarrow -i|b'\>$, driving only the $|a'\>|0\> \rightarrow
|a\>|1_a\>$ transition by choosing $\Omega_b=0$ in $U_{\rm sel}^p$
leads to \be |\Psi_{n-1}\> \rightarrow \alpha |a,1_a\> |\psi_a\> -
i\beta |b',0_b\> |\psi_b\>,\ee where the cavity mode $b$ remains
empty and level $|b\>$ is not yet populated. Now we transform
$|a\>\rightarrow |b\>$, then apply $U_{\rm sel}^p$ again and,
since the photon in the mode $a$ can not be absorbed, end up with
\be |\Psi_n\>= |b\>\otimes \big(\alpha |1_a\>|\psi_a\>+
\beta|1_b\> |\psi_b\>\big),\ee where the atom decouples in its
final state $|b\>$.

\subsubsection{Adiabatic passage}

\begin{figure}[t]
\begin{center}
\epsfig{file=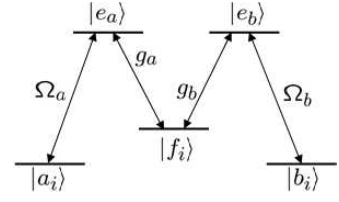,angle=0,width=0.5\linewidth}
\end{center}
\caption {Atomic level structure: level $|a_i\>$ ($|b_i\>$) are
coupled to $|f_i\>$ via an adiabatic passage driven by a laser with
Rabi frequency $\Omega_a$ ($\Omega_b$) and the cavity mode $a$ ($b$)
with coupling strength $g_a$ ($g_b$).} \label{polstd}
\end{figure}

We consider an effective three-level system, with atomic ground
states $|a_i\>$, $|b_i\>$ and $|f_i\>$, which couples to the two
cavity-modes $a$ and $b$ through two independent adiabatic passages
controlled by two external lasers with Rabi frequencies $\Omega_a$
and $\Omega_b$ as depicted in Fig.\ \ref{polstd}.

In each generation step the standard map is achieved, \bea |a_i\>
&\rightarrow& |f_i\>|1_a\> \rightarrow |b_i\> |1_a\>, \nonumber\\
|b_i\> &\rightarrow& |f_i\>|1_b\> \rightarrow |b_i\> |1_b\>, \eea
where we first applied the adiabatic passages and then a unitary
operation $|f_i\>\rightarrow |b_i\>$. In general, we have
$i=1,\dots,D$. Instead of an effective $2D$-level atomic system, as
for the generation of time-bin entangled ${\rm MPS}_D$, we require a
$3D$-level atomic system.

Note that the results for the generation of time-bin qubits apply:
for W-type, GHZ and cluster states we require only $D=1$ and one
additional level, i.e., an effective four-level atom.

\section{Sequential generation of atomic states}\label{sec:AtomicStates}

For the generation of atomic multi-qubit states the cavity mode and
the atom interchange their function. Now, the cavity field is
employed as the ancillary system to interact with
initially uncorrelated atoms which sequentially pass through the
cavity. Therefore, we require a very stable cavity field which is
provided typically by microwave cavities \cite{Haroche2000,
Walther2001}. In the second part of this section, we consider another
scenario where the atomic qubits interact directly via a common
cavity mode \cite{Jauslin2005}.

\subsection{Cavity acts as ancillary system}

\begin{figure}[t]
\begin{center}
\epsfig{file=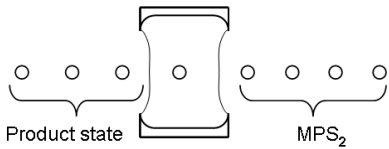,angle=0,width=0.6\linewidth}
\end{center}
\vspace*{-0.5cm}\caption {A stream of uncorrelated atoms crosses a
cavity. The atomic qubits couple sequentially to the cavity mode,
which acts as a $2$-dimensional ancillary system. If an arbitrary
unitary operation can be realized between the cavity qubit and the
atomic qubits, the class of entangled multi-qubit atomic states at
the output are equivalent to the class of ${\rm MPS}_2$.}
\label{cavanc}
\end{figure}

In the first scenario the atoms pass through the cavity in such a
manner that only one atom couples to a single cavity mode at a time,
as depicted in Fig.\ \ref{cavanc}. We employ the same atomic level
configuration as in section 3A and define the ancilla qubit as the
cavity mode Fock states $|0\>$ and $|1\>$. The atom-cavity system is
then described by the selective Hamiltonian from Eq.\ \ref{eq:hsel}.
Choosing the field frequency and pulse length appropriately leads to
an ISWAP gate between the photonic and the atomic qubit. In the
basis $\{|b,0\>, |b,1\>, |a,0\>, |a,1\>\}$, \be {\rm ISWAP} =
\left(\begin{array}{cccc} 1&0&0&0\\0&0&i&0\\0&i&0&0\\0&0&0&1
\end{array}\right).\ee

Three applications of the ISWAP gate accompanied by local unitary
operations suffice for an arbitrary two-qubit gate
\cite{Whaley2004}, though they could only be applied in setups where
the position of the atoms is fully controlled~\cite{Lange2004,
Darquie2005, Nussmann2005}. In fact, it would not only allow the
generation of all MPS with two-dimensional bonds, but also all other
states since the atoms could be moved back and forth, allowing for
universal quantum computing. One of those setups use optical
cavities and, unfortunately,  the cavity decay is not negligible on
the time-scale in which the atoms are moved in and out of the
cavity.

For setups with high-finesse microwave cavities~\cite{Walther2001,
Haroche2000}, the atoms pass once the cavity and local rotations can
be implemented via suitably placed Ramsey zones. So the natural
question arises: which states can be generated if only one ISWAP
gate and local unitaries on the atomic qubits are available?

If all atoms are prepared in state $|a\>$ and the cavity mode is
initially empty, sequential application of the $\sqrt{\rm ISWAP}$
gate leads to \bea |0\> &\rightarrow& \frac{1}{\sqrt{2}} |0\>|a_1\>
+ \frac{i}{\sqrt{2}} |1\> |b_1\> \nonumber\\ &\rightarrow&
\frac{1}{2} |0\>|a_2,a_1\> + \frac{i}{2} |1\> |b_2,a_1\> +
\frac{i}{\sqrt{2}} |1\> |a_2,b_1\> \nonumber\\ &\rightarrow& \dots
\nonumber\\ &\rightarrow& \frac{1}{\sqrt{2^{n-1}}} |0\>|a_{n-1}
\dots a_1\> \nonumber\\&& + \frac{i}{\sqrt{2^{n-1}}} |1\>
|b_{n-1},a_{n-2},\dots ,a_1\> +\dots \nonumber\\&&
+\frac{i}{\sqrt{2}} |1\>|a_{n-1},\dots ,a_2, b_1\>.\eea This W-type
state is still entangled with the cavity qubit. One way of solving
this problem is to measure the cavity qubit in the basis $(|0\>\pm
|1\>)/\sqrt{2}$.

\begin{figure}[t]
\begin{center}
\epsfig{file=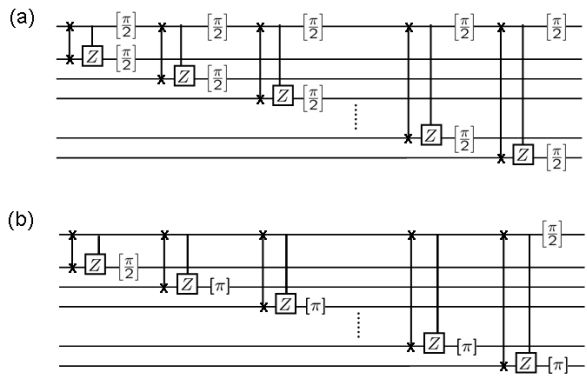,angle=0,width=0.9\linewidth}
\end{center}
\vspace*{-0.5cm}\caption{Cluster state generation. a) An ISWAP gate
is performed sequentially between the atomic qubits and the cavity
mode. b) One can as well assume that the local unitaries are
performed on the atomic qubit $k$ after the $k$th SWAP. Therefore
they can be compensated in a final step.} \label{atcluster}
\end{figure}

Let us concentrate on the generation of the one-dimensional cluster
state given in Eq.~\ref{cluster}. The required control Z (CZ) gate
between neighboring qubits can be realized through a CZ gate between
the ancilla and the first qubit followed by a SWAP operation on the
ancilla and the second qubit. Therefore, we decompose \be {\rm
ISWAP} = \left(
\begin{array}{cccc} 1&0&0&0\\0&0&1&0\\0&1&0&0\\0&0&0&1
\end{array}\right)\cdot
\left(\begin{array}{cccc} 1&0&0&0\\0&i&0&0\\0&0&i&0\\0&0&0&1
\end{array}\right),\ee where the first matrix represents the SWAP
operation and the second matrix is equivalent to the CZ gate up to
local operations. As has been depicted in Fig.\ \ref{atcluster} (a),
the ISWAP can then be written as \be {\rm ISWAP}= i\;{\rm
SWAP}\;{\rm CZ}\;\left[R_z(\pi/2) \otimes R_z(\pi/2)\right], \ee
with \be R_z(\phi)= e^{-i\sigma_z\phi/2}= \left(\begin{array}{cc}
e^{-i\phi/2}&0\\0&e^{i\phi/2}\end{array}\right).\ee

Since it is difficult to perform local operations on the cavity qubit, say
in step $k$, we make use of the SWAP and apply it instead in step
$k+1$ after the SWAP on qubit $k+1$. Fortunately, $R_z(\pi/2)$
commutes with the CZ gate and we obtain the recipe shown in Fig.\
\ref{atcluster} (b). The last unitary on the ancilla can be ignored
because it has no influence on the desired multi-qubit state. In
order to obtain the cluster state as defined in Eq.\ \ref{cluster},
we have to compensate for the local unitaries on the atoms and
prepare all atoms initially in superpositions of $(|0\>+ |1\>)/
\sqrt{2}$.

\subsection{No ancilla: atoms interact via cavity mode}

\begin{figure}[t]
\begin{center}
\epsfig{file=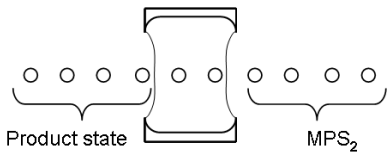,angle=0,width=0.6\linewidth}
\end{center}
\vspace*{-0.5cm}\caption {A stream of uncorrelated atoms passes a
cavity. Two atomic qubits couple to each other via the common cavity
mode. If an arbitrary unitary operation can be realized between
them, the class of entangled multi-qubit atomic states at the output
are quivalent to the class of ${\rm MPS}_2$.} \label{noanc}
\end{figure}

Using the cavity mode as an ancilla has the disadvantage that it may
decay during the time-interval between two successive atoms. This
problem can be avoided by a direct interaction between the atomic
qubits via the common cavity mode. As sketched in Fig.\ \ref{noanc},
there are always two atoms at the same time inside the cavity. Zheng
and Guo~\cite{Zheng2000} proposed the implementation of a $\sqrt{\rm
ISWAP}$ between the atomic qubits via a dispersive scheme. Note that
in order to implement a CNOT an additional atomic level would be
required~\cite{Zheng2000}. Recent experiments~\cite{Nussmann2005,
Lange2004} raise hope that it will soon be possible to move two
atoms into a cavity in a well controlled manner. Then the $\sqrt{\rm
ISWAP}$ accompanied by local unitaries suffices to perform an
arbitray two-qubit operation.

If we consider that both atoms cross the cavity only
once~\cite{Haroche2001}, where an $\sqrt{\rm ISWAP}$ has been
demonstrated, the question arises again: which mutli-qubit atomic
states can be sequentially generated in this manner? Since the
atomic qubits can be manipulated locally before and after the gate
with Ramsey zones, we gain more possibilities than in the
corresponding ancilla case.

\begin{figure}[t]
\begin{center}
\epsfig{file=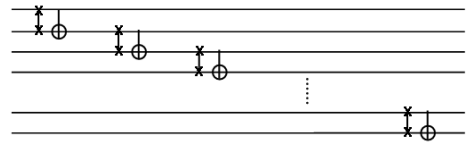,angle=0,width=0.7\linewidth}
\end{center}
\vspace*{-0.5cm}\caption{The ISWAP gate between neighboring qubits
is equivalent to a SWAP followed by a CNOT up to local unitaries.
The latter can be compensated since local operations can be applied
to the atomic qubits with Ramsey zones.} \label{atghz}
\end{figure}

The generation of the W and cluster states follows the lines of the
previous case of secion IVB, since we only replace the ancilla by
the neighbour qubit and the superfluous SWAP gate does not affect
the desired output state. In order to engineer a GHZ state, we use
another decomposition of ISWAP gate, given by \bea {\rm ISWAP}&=&
i\;{\rm SWAP}\;\left[R_z(\pi/2) \otimes
R_z(\pi/2)\right]\nonumber\\&& \times \left[\one \otimes H \right]
{\rm CNOT} \left[\one \otimes H \right], \eea where the Hadamard
gate is given by \be H= \frac{1}{\sqrt{2}} \left(\begin{array}{cc}
1&1\\1&-1\end{array}\right).\ee Since we can compensate the local
unitaries acting on the atomic qubits, we end up with a sequential
application of a CNOT followed by a SWAP on all neighboring qubits
as depicted in Fig.~\ref{atghz}.

Now, let us assume that our initial state $|\psi_I\>$ has all atoms
prepared in level $|a\>$, except the first which enters the cavity
in the superposition $(|a\>+|b\>)/\sqrt{2}$, \bea |\psi_I \>
&\rightarrow& \frac{1}{\sqrt{2}} |a\dots a\>
\otimes (|a,a\>+|b,b \>) \nonumber\\
&\rightarrow& \dots \nonumber\\ &\rightarrow& \frac{1}{\sqrt{2}}
(|a,\dots, a\>+|b, \dots ,b \>).\eea Here, the arrows in step $i$
indicate the application of a CNOT and a SWAP between qubits $i$ and
$i+1$, producing at the end the desired GHZ state.

Further experimental efforts will make use of two consecutive
cavities in the same setup \cite{privatecommunication}. This opens
up new possibilities in terms of state engineering in the light of
the present work. Three cavities in a row, which are crossed by
atoms such that unitary operations may be performed between the
cavities, would even allow for arbitrary two-qubit operations
between the atoms and therefore lead to a class of states equivalent
to the class of ${\rm MPS}_2$.

\section{Conclusions}

We developed a formalism to describe the sequential generation of
entangled multi-qubit states. It classifies the states achievable by a
$D$-dimensional ancilla as MPS with $D$-dimensional bonds and
provides a recipe for the sequential generation of any state. It turns out
that all states that can be generated non-deterministically with a
$D$-dimensional ancilla also belong to the class of MPS with
$D$-dimensional bonds. Therefore, we will be able to provide a recipe
for their deterministic generation as well. Remark that the
formalism applies also to a situation where qubits interact directly
in a sequential manner.

For the generation of entangled multi-qubit photonic states by a
cavity QED single-photon source, we propose am implementation of an
arbitrary interaction between the atom and a cavity qubit defined
either by the absence and the presence of a photon or by the
polarization of single excitation in the cavity. Moreover, we
discussed the case of an adiabatic passage for the photon generation
as being employed for single-photon generation in current cavity QED
experiments \cite{Kuhn2002, McKeever2004, Lange2004}.

For coherent microwave cavity QED experiments \cite{Haroche2000},
where atoms sequentially cross a cavity and interact with the same
cavity mode, we give a recipe for the generation of W-type states as
well as cluster states. We considered also the case of direct
coupling of successive atoms via the cavity mode. For this case we
show how to generate GHZ states, in addition to the W-type  and cluster states.

Other physical scenarios as a light pulse crossing several atomic
ensembles~\cite{Polzik2003} or trapped ion experiments, where each
ion interacts sequentially with a collective mode of the
motion~\cite{NIST,Innsbruck,CiracZoller}, may also be described by
the present formalism.

\begin{acknowledgments}
We would like to thank G. Giedke and F. Verstraete for useful and
stimulating discussions. This work was supported by the European
Union projects COVAQIAL, QAP, CONQUEST and SCALA. C.S. acknowledges
support from the UK Engineering and Physical Sciences Research
Council and partial support from the Quantum Information Processing
Interdisciplinary Research Collaboration (QIP IRC). K.H.
acknowledges support from the Austrian Science Foundation. E.S.
acknowledges financial support from DFG SFB 631 and EU EuroSQIP
project.
\end{acknowledgments}

\end{document}